\documentclass[journal]{IEEEtran}

\IEEEoverridecommandlockouts                              

\usepackage{numbersUncertainty}
\usepackage{enumitem}       

\newcommand{\Tset}{\mathcal{T}}

\title{Fusion framework and multimodality for the Laplacian approximation of Bayesian neural networks  \\
}

\author{Magnus Malmstr{\"o}m,  Isaac Skog \IEEEmembership{Senior Member, IEEE}, Daniel Axehill \IEEEmembership{Senior Member, IEEE}, Fredrik Gustafsson \IEEEmembership{Fellow, IEEE} 
	\thanks{ This work is supported by Sweden's innovation agency, Vinnova, through project iQDeep (project number 2018-02700). }
		\thanks{Magnus Malmstr{\"o}m, Daniel Axehill, and Fredrik Gustafsson, is with Link{\"o}ping University (e-mail: \{magnus.malmstrom, daniel.axehill fredrik.gustafsson\}@liu.se)  }
	\thanks{Isaac Skog, is with Uppsala University  (e-mail:  isaac.skog@angstrom.uu.se)}
}

\begin{document}
\maketitle
 \bstctlcite{IEEEexample:BSTcontrol}

\thispagestyle{empty}
\pagestyle{empty}


\begin{abstract}
    This paper considers the problem of sequential fusion of predictions from neural networks (\nn) and fusion of predictions from multiple \nn. This fusion strategy increases the robustness, i.e., reduces the impact of one incorrect classification and detection of outliers the \nn has not seen during training.
    This paper uses Laplacian approximation of Bayesian \nn{s} (\bnn{s}) to quantify the uncertainty necessary for fusion. Here, an extension is proposed such that the prediction of the \nn can be represented by multimodal distributions. 
     Regarding calibration of the estimated uncertainty in the prediction, the performance is significantly improved by having the flexibility to represent a multimodal distribution. Two class classical image classification tasks, i.e., \mnist and \cfarTen, and image sequences from camera traps of carnivores in Swedish forests have been used to demonstrate the fusion strategies and proposed extension to the Laplacian approximation.

\end{abstract}



\section{Introduction}
This paper studies how to fuse the prediction from multiple neural networks (\nn) classifiers.
Two problem scenarios are considered. Firstly, the problem of fusing the predictions from multiple \nn classifiers that attempt to classify the same object. Secondly, to fuse the predictions from a single \nn classifier that are given a sequence of inputs known to belong to the same class. 

One often uses multiple algorithms and methods that work in parallel to have redundancy in a decision process. The predictions are then combined to make the decision more robust. However, without knowledge of the uncertainty in the prediction, it is unclear how the different predictions should be weighted. Hence, it is necessary to have good knowledge regarding the uncertainty in the prediction from the model. 

In recent years, \nn{s} have had great success generating images from text \cite{ramesh2021zero}, mastering board games such as GO \cite{silver2016mastering}, and in various control tasks \cite{li2017deep,zhang2019safe}. Despite their tremendous success, there is still limited use of \nn{s} in safety-critical applications, e.g., medical imaging and autonomous driving \cite{paleyes2020challenges,bagloee2016autonomous,grigorescu2020survey}.
One of the main reasons \nn{s} have not revolutionized autonomous vehicles yet is the lack of knowledge of the uncertainty in their predictions. 
Take, for example, the infamous accident in 2018 by one of Uber's autonomous vehicles \cite{Uber}. Here, the lack of reliable classification uncertainty by the \nn of the surrounding object was a contributing factor to the fatal outcome of the accident. 
To create a more robust decision, the fusion of the prediction from sensors such as cameras and lidars could have been helpful in this scenario.

The problem of quantifying the uncertainty in the prediction of \nn has lately been gaining interest with many suggested methods \cite{damour2020underspecification, Ghahramani2015,patel2022accurate,ovadia2019can,lin2022uncertainty,izmailov2021dangers}. See \cite{gawlikowski2021survey} for a survey over different methods. 
Broadly, one can separate these methods to quantify uncertainty in the prediction into two categories. The first category of methods is based on designing the structure of the \nn such that it learns its own uncertainty \cite{charpentier2020posterior, Kendall2016,gustafsson2019dctd,eldesokey2020uncertainty}. The second category is based on creating an ensemble of predictions, from which the uncertainty in the prediction could be computed \cite{Blundell2015,lakshminarayanan2016simple, Gal2015a, Taye2018,maddox2019simple,osawa2019practical,ayhan2018test,ilg2018uncertainty,carannante2021enhanced,malmstrom2023uncertainty,immer2021improving}.  

This paper will consider methods from the second category. 
The naive approach would be to independently train multiple \nn on the same task \cite{lakshminarayanan2016simple}. This method is frequently referred to as \textit{deep ensemble}. The major disadvantage of this approach is that it is very computationally costly since it requires training multiple \nn where only training one takes a lot of computational resources. 
The so-called test time augmentation is another method to create the ensemble. Here, a single \nn creates an ensemble of predictions by predicting the slightly modified version of the input \cite{ayhan2018test}. Test time augmentation is typical in medical image classification, where little data is available. Here, the modification could, for example, be to rotate the image. However, this method does not consider the correlation from using the same \nn to predict the output of all the augmented inputs.

Instead of assuming the parameters of the \nn to be fixed, one can assume that they have some distribution from which one can draw samples to create the ensemble, i.e.,  a Bayesian \nn (\bnn) \cite{Bishop2006}.
However, to train a \bnn is not straightforward and can be computationally expensive, and hence, one is often referred to use some approximation method. For example, values of the parameters could be sampled during the later part of the training \cite{maddox2019simple,osawa2019practical}, or the distribution can be assumed to be given by some already used regularization techniques from which samples can be drawn during inference \cite{Gal2015a, Taye2018}. 
Another well-used approximation is the so-called Laplacian approximation of a \bnn, where the curvature of the likelihood function gives the distribution of the parameters \cite{immer2021improving,martens2015optimizing,kristiadi2020being, Bishop2006,mackay1992practical}. 
In particular, this paper will focus on the linearized Laplacian approximation (\lla) \cite{malmstrom2023uncertainty,deng2022accelerated}, where the so-called \textit{delta method} is used to model the distribution for the prediction rather than the distribution for the parameters \cite{Hannelore2011, Gene1997,malmstrom2021}.  

A shortcoming with Laplacian approximation is that it is a local approach that only quantifies the uncertainty in a neighborhood of some given parameters. Hence, it lacks the expressiveness to represent a multimodal distribution. To solve this shortcoming,  we propose the ensemble \lla (\ella). The proposed method combines deep-ensemble and Laplacian approximation, training multiple \nn{s} making a Laplacian approximation for each of the \nn{s}.

The contribution of this paper is three-fold. Firstly, the paper presents a method to fuse predictions from multiple classifications, which achieves a  more accurate classifier. Both for the case using multiple classifiers and when the classification comes from a sequence of inputs known to belong to the same class. Here, fusing the information is shown to lead to more robust decisions.
Secondly, the delta method for the Laplacian approximation of \bnn presented in \cite{malmstrom2023uncertainty} is extended such that the estimated probability can represent multimodal distributions.
Thirdly, the presented methods demonstrate efficiency in detecting out-of-distribution examples. 

\section{Fusion of Classifiers}
We consider the problem of fusing several classifiers' predicted probability mass functions (\pmf) from a Bayesian perspective. Let $y\in\{1,\ldots,M\}$ and $x\in\mathbb{R}^{n_x}$ denote the class label and input, respectively. Further, assume that the joint probability distribution $p(X, Y)$ can describe the underlying data-generating process. There are $C$ training data sets generated as 
\begin{align} 
  \begin{split}
     \mathcal{T}^{(c)}  =  \{x_i, y_i \}_{i=1}^{N_c},\quad (x_i,y_i)\stackrel{\text{\tiny iid}}{\sim}p(X,Y), \quad c=1,2,\!\dots,C.
  \end{split}
\end{align}
available to train the classifiers. During the training-phase, the classifiers try to learn (identify) a function $f(x|\mathcal{T}^{(c)})$ that approximates the conditional distribution $p(Y=y|X=x)$ from the training data $\mathcal{T}^{(c)}$. That is, after the classifiers have been trained, we have the \pmf approximations 
\begin{align}
  p(Y=y|X=x) \approx f(x|\mathcal{T}^{(c)}),\quad c=1,2,\dots,C.
\end{align}
In the classification phase, $L$ inputs (assumed to be known to belong to the same class) are generated as $x^\star_l\sim p(X, Y=y^\star)$ for $l=1,2,\dots, L$. Inserting the new inputs into the classifiers yields the \pmf estimates
\begin{align} \label{eq:individual_pmf}
  \hat{p}_{lc}(y^\star) = f(x^\star_l|\mathcal{T}^{(c)}), \quad l=1,\dots L, \ c=1,\dots,C.
\end{align}
The fundamental question is how to fuse the estimates $\hat{p}_{lc}(y)$ given that we know that all $x^\star_l$ are input samples corresponding to the same, but unknown class $y^\star$? That is, to estimate
\begin{align}  \label{eq:conditional_pmf}
  p(y^{\star}|x_{1:L}^{\star}) \triangleq  p(y^{\star} |x_{1:L}^{\star}, \Tset^{(1)}, \! \! \ldots, \! \Tset^{(C)}).
\end{align}

As a remark, most practical use cases have either $L=1$ (apply several classifiers to one input) or $C=1$ (apply the same classifier to several images, e.g., from a video stream), but both cases and their generalization will be treated in parallel in the sequel. 

\subsection{Fusion using non-parametric models}
If all inputs and training data sets are independent, then the Bayes rule gives that
\begin{align}
  p(y|x_{1:L}^{\star})\propto  \prod_{l=1}^L\prod_{c=1}^C p_{lc}(y).
\end{align}
Thus, one reasonable way to fuse the \pmf estimates is 
\begin{align}
  \hat{p}(y|x_{1:L}^{\star})=\frac{\prod_{l=1}^L\prod_{c=1}^C \hat{p}_{lc}(y)}{\sum_{m=1}^{M} \prod_{l=1}^L\prod_{c=1}^C \hat{p}_{lc}(m)}.
\end{align}
If the relative quality (uncertainty) of the estimates $\hat{p}_{lc}(y)$ are known and represented by the scalar weights $w_{lc}$, $\sum_{lc}w_{lc}=1$, they may also be fused as
\begin{align} \label{eq:fusion_weight}
  \hat{p}(y|x_{1:L}^{\star}) \propto \prod_{l=1}^L \prod_{c=1}^C  \hat{p}_{lc}^{w_{lc}}(y).
\end{align} 
Note that this fusion rule is equivalent to log-linear pooling \cite{koliander2022fusion}. Compared to our Bayesian approach, this method is rather {\em ad-hoc}, and the weights $w_{lc}$ reflect reliability in each classifier rather than a stochastic measure of the expected performance. Furthermore, the weights do not depend on the input, so the relative strength of each classifier in different regions is not exploited.

\subsection{Fusion using parametric models}
So far, we have assumed a black-box structure of the classifier, which only depends on the training data set $\hat{p}_{lc}(y) = f(x^\star_l|\mathcal{T}^{(c)})$. This assumption covers simple classifiers such as the nearest neighbor. In the sequel, we will consider parametric functions, such as \nn{s}, where the classifier explicitly depends on a parameter $\theta$, which is estimated as $\hat{\theta}^c_N$ from the training data $\mathcal{T}^{(c)}$. This dependence will be made explicit in the sequel, so we denote the conditional probabilities
\begin{align}
  \hat{p}_{lc}(y^\star) \triangleq  f(x^\star_l|\hat{\theta}^c_N), \quad \forall l,c.
\end{align}
The uncertainty in $\hat{\theta}^c_N$ implies that the conditional probability $f(x^\star_l|\hat{\theta}^c_N)$ is itself a distribution. This fact is the key point with our approach that allows the fusion of classifiers where the reliability of each one may depend on both the input and the quality of the training data, as reflected in the uncertainty of the parameter estimate. A requirement to be able to do this is to specify the posterior distribution of the parameters $p(\theta|\mathcal{T}^{(c)})$.

Now, a distribution of distributions is quite a complex object to handle, where no analytical expressions can be expected. One remedy is to represent the distribution of the estimated parameters $\hat{\theta}^c_N$ with Monte Carlo (\mc) samples $\theta^{c(k)}$. That is, we use the approximation
\begin{subequations} \label{eq:mc_samp_fg_prod}
\begin{align} 
\theta^{c(k)} &\sim p(\theta|\mathcal{T}^{(c)}), \quad k=1,2\dots, K, \\
  \hat{p}_{lc}(y^\star) &\approx  \frac{1}{K} \sum_{k=1}^K f(x^\star_l|\theta^{c(k)}), \quad \forall l,c, \\
 \hat{p}(y^\star |x_{1:L}^{\star}) & \propto \prod_{l=1}^L \prod_{c=1}^{C} 
 \hat{p}_{lc}(y^\star).  \label{eq:mc_samp_fg_prod_fusion}
\end{align}
\end{subequations}
Here, $K$ denotes the number of samples used in the \mc sampling.
One major shortcoming with this formulation is that every classifier is represented with only a point estimate, which does not utilize the fact that we are dealing with a distribution of distributions. This is because the fusion is done after the estimation using the \mc samples. A remedy for this problem is to perform the fusion before the sampling stage. However, quantifying $p(\theta|\mathcal{T}^{(c)})$ is a difficult problem, hence also to fuse the predictions from all the classifiers, both over the classifiers and also over the sequence of inputs. 

Another shortcoming with approximating $p(y^\star|x^\star_{1:L})$ by \mc sampling from \eqref{eq:mc_samp_fg_prod} is that it comes with a high computational cost. This cost is a result from the approximation requiring drawing samples from a high dimensional Gaussian distribution and evaluating the whole \nn multiple times. The high computational complexity is particularly true for \nn classifiers, which have a huge dimension of the parameter space. In \cite{malmstrom2023uncertainty}, a remedy to this is presented, which reduces the dimension in the sampling to an $M$-dimensional space. In this paper, it is done using the so-called delta method \cite{malmstrom2021, Hannelore2011}. Then, the predicted values before the normalization are sampled instead of the values of the parameters $\theta$ of the \nn. Hence, no forward passes of the \nn are required, and the dimension of the distribution from which the samples are drawn is significantly smaller. Apart from decreasing the computational complexity, this might also lead to a more straightforward fusion strategy compared to \eqref{eq:mc_samp_fg_prod}. This method to propagate the uncerainty will further be described in the following sections.

\section{Neural Network Classifiers} 
This section will outline some basic concepts that are required for the two presented methods in this paper to approximate the conditional \pmf given a set of classifiers, i.e., to estimate $\hat{p}(y^{\star}|x^\star_{1:L})$. In this way, uncertainty in the parameters from the estimation step can be preserved and transformed to, first, the output from the last layers of the {\nn}s and, second, to the class probabilities. The goal is to estimate the probability distribution of the class probabilities, including cross-correlations. 

\subsection{The Softmax Operator}
It will be assumed that the \nn classifier has a softmax function in the output layer. This ensures that the model $f(x|\theta^{(c)})$ fulfills the properties associated with a \pmf, i.e., $f_m(x|\theta^{(c)})\geq 0$ $\forall m$ and $\sum_m f_m(x|\theta^{(c)})=1$. Thus, it is assumed that the \nn is structured as
\begin{subequations}
  \begin{equation}\label{eq:basic_model_fm}
  f(x|\theta^{(c)})=\text{softmax}\left(g(x|\theta^{(c)})\right)
  \end{equation}
  where
  \begin{equation}\label{eq:softmax_fm}
  \text{softmax}(z)\triangleq \frac{1}{\sum_{m=1}^{M} e^{z_m}}\begin{bmatrix}
  e^{z_1} \\
  \vdots \\
  e^{z_M}
  \end{bmatrix}.
  \end{equation}
\end{subequations}
  Here the family of functions $g(x|\theta)$ is unconstrained, while the softmax function maps these functions onto the interval $[0,1]$.
 
It should here be noted that the softmax function is invariant to translations, so $\text{softmax}(z)=\text{softmax}(z+\alpha)$ for all $\alpha$. Here, a transformation is needed to make different classifiers comparable. The chosen translation is arbitrary, but a natural choice is
\begin{equation} \label{eq:shifted_softmax}
\bar{g}(x|\theta^{(c)}) \triangleq g(x|\theta^{(c)}) - \max_k g_k(x|\theta^{(c)}) \leq 0.
\end{equation}
This is a sound choice from a numerical point of view since the exponential functions will operate on numbers smaller than zero, and there is no risk of numerical overflow. However, our main reason for transformation is to enable fusion.

\subsection{Monte Carlo Sampling of Class Probabilities}
Define $z\in \mathbb{R}^M$ as the value of the classifiers before the normalization done by the softmax function. Assume the distribution $p(z^{\star}|x^\star_{1:L})$ has been quantified, i.e., a distribution for the fusion of the classifiers before the normalization. 
By combining the \mc sampling technique in \eqref{eq:mc_samp_fg_prod}, we get an \mc approximation of the class probabilities in the following way:
\begin{subequations} \label{eq:pmf2find}
\begin{align}
z^{(k)} & \sim p(z^{\star}|x^\star_{1:L}), \quad k=1,2,\dots,K, \label{eq:pmf2find_sample} \\
\hat{p}^{(k)} &= \mathrm{softmax} \big(z^{(k)}\big),\\
\hat{p}(y^\star|x^\star_{1:L}) &= \frac{1}{K} \sum_{k=1}^K \hat{p}^{(k)}.
\end{align}
\end{subequations}
The uncertainty of the estimated class labels $\hat{p}(y^\star|x^\star_{1:L})$ is here explicitly represented by the point cloud $\hat{p}^{(k)}$. From this, various probabilities can be computed, e.g., the probability that $\hat{p}(y^\star|x^\star_{1:L})$ is true (count the fraction of sample vectors $\hat{p}^{(k)}$ whose $c$'th element is the largest one), or the risk that we decide $c_1$ while $c_2\neq c_1$ is true.

Hence, we have laid the ground for the proposed Bayesian approach to estimate the class probability distribution.
The remaining task is to estimate $p(z^{\star}|x^\star_{1:L})$, preferably without adding a substantial amount of computations in the training and classification stages.

\section{Uncertainty quantification} 
Next, a method to quantify  $p_{lc}(z^\star)$ for \nn{s} is described. After that, the presentation of an extension where the correlation between classifiers is included. This extension is required to estimate $p(z^\star|x^\star_{1:L}) $.

\subsection{Laplacian approximation and the delta method}
Representing $p(\theta|\mathcal{T}^{(c)})$ is challenging.  
Hence, one often has to rely on approximations to represent $p(\theta|\mathcal{T}^{(c)})$. One such approximation is the  Laplacian approximation of \bnn{s} \cite{Bishop2006,mackay1992practical}. It is a local linear approach, that approximates the uncertainty in the parameters using the curvature of the likelihood function around the estimated parameters $\hat{\theta}_N^c$.
Assume some prior of the parameters $p(\theta) = \mathcal{N}(\theta; 0, P_0)$. Then the Laplacian approximation of the posterior $p(\theta| \mathcal{T}^{(c)})$ yields \cite{Bishop2006}  
\begin{subequations} \label{eq:Laplace_fm}
  \begin{align} 
p(\theta|\mathcal{T}^{(c)}) & = \mathcal{N}(\theta; \hat{\theta}_N^c, P_N^{\theta,c}), \label{eq:posteriorLapalce} \\ 
P_N^{\theta,c} & = \bigg(- \frac{\partial^2 L_N(\theta)}{\partial \theta^2}\bigg|_{\theta = \hat{\theta}_N^c} + P_0^{-1}\bigg)^{-1} \label{eq:lapalceParamCov}.
\end{align}
\end{subequations}
Here 
\begin{equation} \label{eq:crossentroAll_fm}
L_N(\theta)= \sum_{n=1}^N  \ln f_{y_n}(x_n|\theta)
\end{equation}
denotes the cross-entropy likelihood function~\cite{lindholm2022machine} where $y_n$ is used as an index operator for the subscript $m$ of $f_m(x|\theta)$.
The maximum a posteriori estimate of the parameter $\hat{\theta}_N$ is given by
\begin{equation} \label{eq:map_estimate_fm}
\hat{\theta}_N^c =\argmax_{\theta} p(\theta|\mathcal{T}^{(c)})=\argmax_{\theta} L_N(\theta)+ \ln p(\theta),
\end{equation}
The choice of the prior distribution $p(\theta)= \mathcal{N}(\theta; 0, P_0)$, corresponds to the so-called $L2$-regularization.

Under the assumption that the true model belongs to the considered model set, Bernstein-von Mises theorem \cite{Johnstone2010} gives us that the distribution of the maximum posterior estimate asymptotically coincides with the Laplacian approximation in \eqref{eq:Laplace_fm}. Furthermore, the second derivative of the likelihood function approximates the inverse of the Fisher information matrix. In \cite{malmstrom2023uncertainty}, it was shown that for a classification problem, the Fisher information is given by
\begin{subequations} \label{eq:information_with_M_fm}
    \begin{equation}
    \mathcal{I}^{\theta,c} \simeq \sum_{n=1}^{N} \sum_{m=1}^{M}  \eta_{m,n}^c  \frac{\partial g_m \! (x_n|\theta)}{\partial \theta}\Bigg|_{\theta = \hat{\theta}_N^c} \bigg( \! \frac{\partial g_m \!(x_n|\theta)}{\partial \theta} \Bigg|_{\theta = \hat{\theta}_N^c}\! \bigg)^{\! \! \top \!} 
    \end{equation}
    where
    \begin{equation}\label{eq:eta_fm}
    \eta_{m,n}^c \triangleq f_m(x_n|\hat{\theta}_N^c)(1-f_m(x_n|\hat{\theta}_N^c)).
    \end{equation}
\end{subequations}

The delta method relies on linearizing a nonlinear model $g(x^\star_l|\hat{\theta}_N^c)$ to propagate the uncertainty in the parameters to uncertainty in the prediction. 
Using the delta method
\begin{subequations}\label{eq:linearize_fm}
    \begin{equation}
    p_{lc}(z^\star)\approx\mathcal{N}\big(z;\hat{g}_{N}^{lc}, P_N^{g,lc})
    \end{equation}
    where
    \begin{equation} \label{eq:biased_before_norem}
    \hat{g}_{N}^{lc} =\text{E}\{z\}\simeq g(x^\star_l|\hat{\theta}_N^c)
    \end{equation}
    and
    \begin{equation} \label{eq:linearizeCov}
    \begin{split}
    P_N^{g,lc} &= \Cov\{z\}\\
    &\simeq\bigg(\frac{\partial}{\partial \theta} g(x^\star_l|\theta)\big|_{\theta=\hat{\theta}_N^c}\bigg)^\top \! P^{\theta, lc}_N \frac{\partial}{\partial \theta}g(x^\star_l| \theta)\big|_{\theta=\hat{\theta}_N^c}.
    \end{split}
    \end{equation}
\end{subequations}
Since the covariance is invariant to translations, the transformed functions in \eqref{eq:shifted_softmax} also have a Gaussian distribution
\begin{subequations} \label{eq:lla_for_one}
  \begin{equation}
  p_{lc}(z^\star)\approx \mathcal{N}\big(z;\hat{\bar{g}}_{N}^{lc},P_N^{g,lc}\big), 
\end{equation}
where 
\begin{align}
  \hat{\bar{g}}_{N}^{lc} \simeq \bar{g}(x^\star_l|\hat{\theta}_N^c).
\end{align}
\end{subequations}

To summarize, the complete method to quantify uncertainty in the prediction for one classifier is based on two linearizations. One to approximate the uncertainty in the parameters, i.e., the Laplacian approximation in \eqref{eq:lapalceParamCov}, and one to propagate the uncertainty to the prediction, i.e., the delta method in \eqref{eq:linearize_fm}.   
The complete method is here referred to as the \lla.

\subsection{Approximating the covariance}
An \nn often has millions of parameters, so it can be intractable to represent the covariance $P^{\theta,c}_N$ for all the parameters. For example, the covariance can be too large to be stored in the computer's local memory. Hence, one often has to rely on approximations of the covariance, e.g., neglecting some correlation between parameters \cite{martens2015optimizing}, assuming some parameters to be fixed \cite{kristiadi2020being}, or neglecting all the cross-correlation between the parameters (mean-field variational inference) \cite{Blundell2015}. In this work, we follow the second approach where it will be assumed that the parameters of the earlier part of the \nn are fixed and only the parameters in the later layers are assumed to be learned \cite{kristiadi2020being,malmstrom2022detection,malmstrom2023uncertainty}. 
This approximation might be more or less accurate depending on the number of included layers. A scaling of $P^{\theta,c}_N$ with factor $T_{\theta} \geq 1$ can be introduced to compensate somewhat for the approximation error. The scaling can be estimated from validation data as is typically done for temperature scaling \cite{guo2017calibration}.

\subsection{Multiple classifiers}
There might be correlations between the classifications when using a single classifier to classify a set of inputs $x^{\star}_l$, $l=1,\ldots, L$, known to belong to the same class $y^\star$. This correlation can be taken into account using    
\begin{align}
\hat{p}_c(\zeta^c| x_{1:L}^\star)  \approx \mathcal{N} \big( \zeta^c(x^\star); \mathbf{\hat{z}}^c , R_c \big)
\end{align}
where $\zeta^c \in\mathbb{R}^{LM}$ denotes the classifications from the $c$'th classifier for the $L$ different inputs   
\begin{align}
\hat{\zeta}^c = \begin{bmatrix} \hat{\bar{g}}_{N}^{1c} \\ \vdots \\ \hat{\bar{g}}_{N}^{Lc} \end{bmatrix}
\end{align}
and  the block $[R]_{i,j}\in\mathbb{R}^{M,M}$, $i,j=1,\ldots,L$, of the covariance matrix is given by
\begin{equation}
[R_c]_{i,j}=\frac{\partial}{\partial \theta} g(x^\star_i|\theta)^\top\big|_{\theta=\hat{\theta}_N^c} \! P^{\theta}_N \frac{\partial}{\partial \theta}g(x^\star_j| \theta)\big|_{\theta=\hat{\theta}_N^c}.
\end{equation}
Extending the description by using multiple classifiers to classify the sequence as 
\begin{subequations} \label{eq:full_description_seq_multi}
    \begin{align} 
    p(\zeta|x_{1:L}^\star) \approx  \mathcal{N}(\zeta;\mathbf{\hat{\zeta}} ,R)
    \end{align}
    where
    \begin{align} 
    \hat{\zeta}  =  \begin{bmatrix} \mathbf{\hat{\zeta}}^{1} \\ \vdots \\ \mathbf{\hat{\zeta}}^{C}, \end{bmatrix} \in\mathbb{R}^{CLM}, \quad
    R = \begin{bmatrix} R_{1} & \ldots & R_{1,C}   \\ \vdots & \ddots & \vdots\\ R_{C,1} & \ldots & R_{C}    \end{bmatrix} .
    \end{align}
\end{subequations}
Here the block $R_{i,j} \in \mathbb{R}^{LM, LM}$ encodes the cross-correlation between the classifier $i$ and classifier $j$ for all inputs in the sequence. This cross-correlation could, e.g., be caused by using the same data set to train the different classifiers. It is, however, unclear how to compute this cross-correlation between classifiers. 

From aggregated state $\zeta$, the prediction for the different classifiers can be recovered as
\begin{subequations} \label{eq:fusion_prud_is}
  \begin{align}
  & \zeta^{(k)}  \sim \mathcal{N}(\zeta;\hat{\zeta},R), \quad k =1,2,\dots K  \label{eq:fusion_prud_is_sample} \\
  & z_{lc}^{(k)} =  W_{lc} \zeta^{(k)}, \quad \forall l,c
  \end{align}
  where $ W_{lc}$ is used to find the classification of the $c$'th calasssifyer of the $l$'th input, i.e., 
  \begin{equation}
  W_{lc}=\begin{bmatrix}
  0 & 
  \ldots & 0 & I_M & 0 & \ldots & 0   
  \end{bmatrix}\in\mathbb{R}^{M,CLM}.
  \end{equation}
\end{subequations}
Then \mc samples can be used to compute a point estimate of the \pmf as 
\begin{align}
  \hat{p}_{lc}(y^\star) = \frac{1}{K} \sum_{k=1}^{K} \text{softmax}(z_{lc})
\end{align}
Afterward, the fusion formula \eqref{eq:mc_samp_fg_prod_fusion} can be applied to fuse prediction. However, this is still a point estimate, and the fusion does not take that some classifiers might have a larger covariance into consideration. Hence, two strategies for aggregating the extended state of classifiers $\zeta$ that can utilize the information that some classifiers might have a larger covariance than others are presented in the following sections.

%

\section{Fusion of Neural Network Classifiers} \label{sec:fusion}
It is possible to use the information from one prediction from an \nn and fuse it with a prediction from other sources of information (which might also be an \nn) having access to the uncertainty in the prediction for an \nn. 
Given the extended description in \eqref{eq:full_description_seq_multi}, for a set of classifications coming from $C$ classifiers and a sequence of $L$ inputs known to belong to the same class $y^\star$, these predictions can be fused as follows
\begin{subequations} \label{eq:fusion_seq}
    \begin{align}
    P_{N}^{g} &= (H^\top R^{-1} H)^{-1},\\
    \hat{\bar{g}}_{N} &=P_{N}^{g} H^\top R^{-1} \hat{\zeta}
    \end{align}
    where
    \begin{equation}
    H=\begin{bmatrix}
    I_{M} \\
    \vdots \\
    I_{M}
    \end{bmatrix}\in\mathbb{R}^{CLM ,M}.
    \end{equation}
\end{subequations}
Due to the Gaussian approximation in \eqref{eq:full_description_seq_multi}, the fused information will also be Gaussian distributed. Hence, the distribution before the normalization can be approximated as 
\begin{align} \label{eq:fusion_dist_fm}
  p(z^\star|x^\star_{1:L})  \approx \mathcal{N}\big(z; \bar{g}_N, \bar{P}_N^{g}),
\end{align}
where \mc samples as in \eqref{eq:pmf2find} can be applied to estimate the desired \pmf. That is, by changing \eqref{eq:pmf2find_sample} to
  \begin{align} 
  &z^{(k)} \sim \mathcal{N}\big(z; \hat{\bar{g}}_{N}, P_N^{g}), \quad k \!=\!1,2,\dots K. \label{eq:gmarg_g_fusion}
  \end{align}
%

In this paper, two special cases of \eqref{eq:full_description_seq_multi} are investigated:
\begin{enumerate}[label=(\roman*)]
    \item Multiple \nn independently train and classify the same input, i.e., $L=1$.
    \item An \nn is used to classify a sequence of inputs known to belong to the same class, i.e., $C=1$. 
\end{enumerate}
These are the edge cases, where a combination results in the general case. 
The first case is motivated by safety-critical systems. There is often more than one sensor measuring the environment. Multiple sensors measuring the same thing are often used to add redundancy to the system. 

For the second case, if one classification is uncertain, the additional information that all the inputs in the sequence belong to the same class will result in a more robust classification. 
In a standard object detection algorithm, this setup can be used on cropped images of the tracked object. After the object detection algorithm finds a bounding box locating an object, it crops the image and sends it to a new classifier that classifies the cropped image. Then, the sequential fusion can be used on the cropped images. 
The second case of fusion strategies can be used in the case of test time augmentation where a single \nn is predicting the output from multiple augmented versions of the same input \cite{ayhan2018test}.

\begin{figure*}[tb!]
	\centering
\scalebox{0.65}{
  \begin{tikzpicture}[scale=0.5,node distance=2cm]
    \node[inner sep=25pt] (start) at (0,-5) {};
\node (object_2) [process, right of=start,xshift=1cm, yshift=1.4cm] { \textcolor{white}{Transform the output before the softmax step using \eqref{eq:shifted_softmax}. } }; 
\node (filter_2) [process, right of=object_2,xshift=3cm] {\textcolor{white}{Estimate $p_{lc}(z^\star)$ using \protect{\lla} in \eqref{eq:lla_for_one}.} };
\draw [arrow] (object_2) --node[anchor=north] {}   (filter_2);

\node (object_1) [process, right of=start,xshift=1.5cm, yshift=0.7cm] { \textcolor{white}{Transform the output before the softmax step using \eqref{eq:shifted_softmax}. } }; 
\node (filter_1) [process, right of=object_1,xshift=3cm] {\textcolor{white}{Estimate $p_{lc}(z^\star)$ using \protect{\lla} in \eqref{eq:lla_for_one}.} };
\draw [arrow] (object_1) --node[anchor=north] {}   (filter_1);

\node (robust) [process, right of=filter_1, xshift=3cm] { Aggregate the classifiers using \eqref{eq:full_description_seq_multi}.};

\draw [arrow] (filter_1) --node[anchor=north] {}   (robust);
\draw [arrow] (filter_2) --node[anchor=north] {}   (robust);
\node (filter_1) [process, right of=object_1,xshift=3cm] {\textcolor{white}{Estimate $p_{lc}(z^\star)$ using \protect{\lla} in \eqref{eq:lla_for_one}.} };

\node (object) [process, right of=start,xshift=2cm] {Transform the output before the softmax step using \eqref{eq:shifted_softmax}. }; 
\node (filter) [process, right of=object,xshift=3cm] {Estimate $p_{lc}(z^\star)$ using \protect{\lla} in \eqref{eq:lla_for_one}. };

\node (fusion) [process, right of=robust, xshift=2.5cm] { Compress $\zeta$ using the fusion in \eqref{eq:fusion_seq} or \protect{\ella.}};

\node (pmf_est) [process, right of=fusion, xshift=3cm] { Estimate the \protect{\pmf} by \protect{\mc} sampling using \eqref{eq:pmf2find}.};

\draw [arrow] (start.east) -- node[anchor=north] {$x_l, \hat{\theta}_N^{c}$} (object);
\coordinate[above of = start , xshift=0.5cm, yshift=-1.3cm, name = start_1] {};
\draw [arrow] (start_1.east) -- node[anchor=north] {}  (object_1);
\coordinate[above of = start , xshift=0cm, yshift=-0.6cm, name = start_2] {};
\draw [arrow] (start_2.east) -- node[anchor=north] {}  (object_2);

\draw [arrow] (object) --node[anchor=north] {$\bar{g}(x_l|\hat{\theta}^{c}_N)$}   (filter);

\draw [arrow] (filter) -- node[anchor=north] {$p_{lc}(z^\star)$}  (robust);
\draw [arrow] (robust) -- node[anchor=north] {$\zeta$}  (fusion);
\draw [arrow] (fusion) -- node[anchor=north] {$p(z^\star|x_{1:L}^\star)$}  (pmf_est);

\coordinate[right of = pmf_est ,xshift=2cm, name = empty_out] {};
\draw [arrow] (pmf_est) -- node[anchor=north] {$\hat{p}(y^\star|x_{1:L}^\star)$} (empty_out);


\end{tikzpicture}

  }
\caption{ A block diagram of the proposed strategy to estimate $\hat{p}(y^\star|x_{1:L}^\star)$. Note that, two different methods are presented in this paper to compress the aggregated classification to quantify $p(z^\star|x_{1:L}^\star)$. Namely, the fusion strategy proposed in \cref{sec:fusion} or the \protect{\ella} proposed in \cref{sec:ensablemLinearization}.   }
  \label{fig:block_diagram_fm}
\end{figure*}
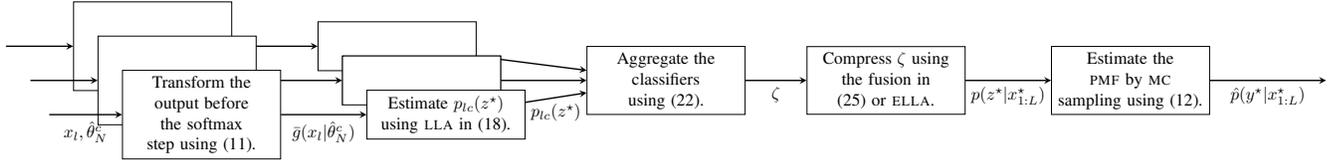

\section{Ensemble linearized Laplacian approximation } \label{sec:ensablemLinearization}
Fusing the prediction using \eqref{eq:fusion_seq} to estimate $p(z^\star|x^\star_{1:L})$ summarizes the information from multiple distributions with a single mode into a new distribution with a single mode. 
The \lla provides a local description of the uncertainty in the prediction, from which efficient \mc sampling can estimate the \pmf, marginalizing the uncertainty in the parameters. 
However, the Laplacian approximation is a local approach requiring more flexibility to represent multimodal distributions. Hence, if the loss of \eqref{eq:crossentroAll_fm} is multimodal, the approximation might be inaccurate. 
On the other hand, the ensemble method, training multiple \nn{s}, e.g., deep ensemble \cite{lakshminarayanan2016simple}, can represent multimodal distributions. However, it is very costly to create a new sample, i.e., since every new sample requires training of a new \nn. 
Hence, we suggest combining the two methods mentioned, i.e.,  training multiple \nn{s} and creating an \lla for each traned \nn. This proposed extension of the \lla is referred to as \ella. 
Instead of drawing samples from the distribution described by \eqref{eq:fusion_dist_fm}, one can draw samples from the full description of the distribution in \eqref{eq:fusion_prud_is}. Afterward, \eqref{eq:pmf2find} can be applied to estimate the desired \pmf. That is, by changing \eqref{eq:pmf2find_sample} to 
\begin{align} \label{eq:multimodality}
  z^{(k)} = \sum_{l=1}^{L} \sum_{c=1}^{C} w_{lc} z_{lc}^{(k)}.
\end{align}
That is, $p(z^\star|x^\star_{1:L}) $ is approximated by $CL$ Gaussian distributions where the parameter is used $w_{lc}$ to weigh the impact of the different modes. For example, could $w_{lc}$  be chosen proportional to the covariance $P_N^{g, lc}$ for the classifier. This results in a method that can represent a multimodal distribution and is efficient to sample from.
The resulting method is similar to a Gaussian mixture model but with some key differences \cite{Bishop2006_gaus_mix}. Instead of independently sampling from the difference modes, the sampling is done over the extended state $\zeta$, where the weighting is done after the sampling. This enables us to include cross-correlation between the modes.  

Similar ideas to combine different methods to quantify the uncertainty have been presented in, e.g.,  \cite{pop2018deep}, where deep ensemble and \mcdropout are combined.
Ideally, each realization should represent a different mode in the distribution. To make this more likely, repulsive training can be used when training the \nn{s} \cite{d2021repulsive}, where it is enforced during the training process that the different \nn{s} converge to different parameters.

The complete strategy proposed in this paper to estimate $\hat{p}(y^\star|x^\star_{1:L})$ is summarized in \cref{fig:block_diagram_fm}. Here, for the compression of $\zeta$, either the fusion strategy in \cref{sec:fusion} or the \ella proposed in this section could be used.

\section{Validation} \label{sec:validation_fm}
Suppose access to some validation data set $\mathcal{V} =\{y^\circ_n,x^\circ_n\}_{n=1}^{N_\circ}$. The question is how to validate that the estimated \pmf resembles the true one. The inherent difficulty is that the validation data, as the training data, consists of inputs with corresponding class labels and not measures of a \pmf. This is one of the reasons why a unified qualitative evaluation metric is lacking for the uncertainty in the prediction \cite{gawlikowski2021survey}. However, some of the most common metrics used are classification accuracy, log-likelihood (\Loglike), prediction entropy, Brier score, expected calibration error (\ece), area under the receiver operating characteristic curve (\auroc), and the area under the precision-recall curve (\aupr) when detecting out-of-distribution samples.

Negative \Loglike, prediction entropy, and Brier score are all proper scoring rules, i.e., they emphasize a careful and honest assessment of the uncertainty \cite{gneiting2007strictly}. However, none of them measure the calibration, i.e., the reliability of the estimated \pmf. Hence, the most important metric is the \ece when comparing the method on in-distribution data. For out-of-distribution data, the \auroc and \aupr are useful metrics, as well as the difference in predicted entropy for in and out-of-distribution data.

\subsection{Accuracy and calibration}
Calculate the $J$-bin histogram defined as
\begin{equation}
B_j =  \bigg \{ n: \frac{j-1}{J} \leq \max_m \hat{p}(m|x^\circ_n) < \frac{j}{J}   \bigg \}
\end{equation}
from the validation data. For a perfect classifier $B_j = \emptyset$ for $j<J$. For a classifier that is just guessing, all sets are of equal size, i.e., $|B_j|=|B_i|$ $\forall i,j$. Note that $\max_m \hat{p}(m|x^\circ_n)\geq 1/M$, so the first bins will be empty if $J>M$.

The accuracy of the classifier is calculated by comparing the size of each set with the actual classification performance within the set. That is,
\begin{subequations}
    \begin{equation}
    \text{acc}(B_j) = \frac{1}{|B_j|} \sum_{n \in B_j} \mathbb{1}\big(\hat{y}^\circ_n = y^\circ_n\big)
    \end{equation}
    where
    \begin{equation}
    \hat{y}^\circ_n=\argmax_m \hat{p}(m|x^\circ_n)
    \end{equation}
\end{subequations}
Instead of certainty, from hereon, the standard and the equivalent notion of confidence will be used \cite{guo2017calibration,vaicenavicius2019evaluating}.
The mean confidence in a set is denoted $\text{conf}(B_j)$ and is defined as
\begin{align}
\text{conf}(B_j) = \frac{1}{|B_j|} \sum_{n \in B_j} \max_m \hat{p}(m|x^\circ_n),
\end{align}
This measures how much the classifier trusts its estimated class labels. In contrast to the accuracy, it does not depend on the annotated class labels $y_n$. Comparing accuracy to confidence gives the \ece, defined as
\begin{align}
\text{\ece} =  \sum_{j}^J \frac{1}{|B_j|} |\text{acc}(B_j)-\text{conf}(B_j)|.
\end{align}
A small value indicates that the weight is a good measure of the actual performance.

\subsection{Log likelihood, predicted entropy, and Brier score}
The total negative \Loglike is given by \eqref{eq:crossentroAll_fm}, while the predicted entropy for one input is given by  
\begin{equation} \label{eq:entropy_fm}
E = -\sum_{m=1}^M  \hat{p}(m|x^\circ_n) \ln \bigg(\hat{p}(m|x^\circ_n) \bigg).
\end{equation}
Here $\hat{p}(m|x^\circ_n)$ denotes some generic estimate of the \pmf.
Hence, the difference between \Loglike and entropy is that to calculate the predicted entropy, the class label $y^\circ$ information is not required. That means the predicted entropy can be computed for out-of-distribution examples. 
For out-of-distribution detections, the difference in entropy between the in-distribution data and out-of-distribution data is of interest, i.e., $\Delta_E = E_{in} -E_{out}$.

The Brier score \cite{gneiting2007strictly,vaicenavicius2019evaluating} corresponds to the least squares fit
\begin{align}
\frac{1}{N_\circ} \sum_{n=1}^{N_\circ}\sum_{m=1}^{M} \bigl(\delta_{m,y_n^\circ}-\hat{p}(m|x^\circ_n)\bigr)^2,
\end{align}
where $\delta_{i,j}$ denotes the Kronecker delta function.

\begin{figure}[tb!]
	\centering
	\includegraphics[trim={6cm 0cm 4cm 0cm},clip,width=0.5\textwidth]{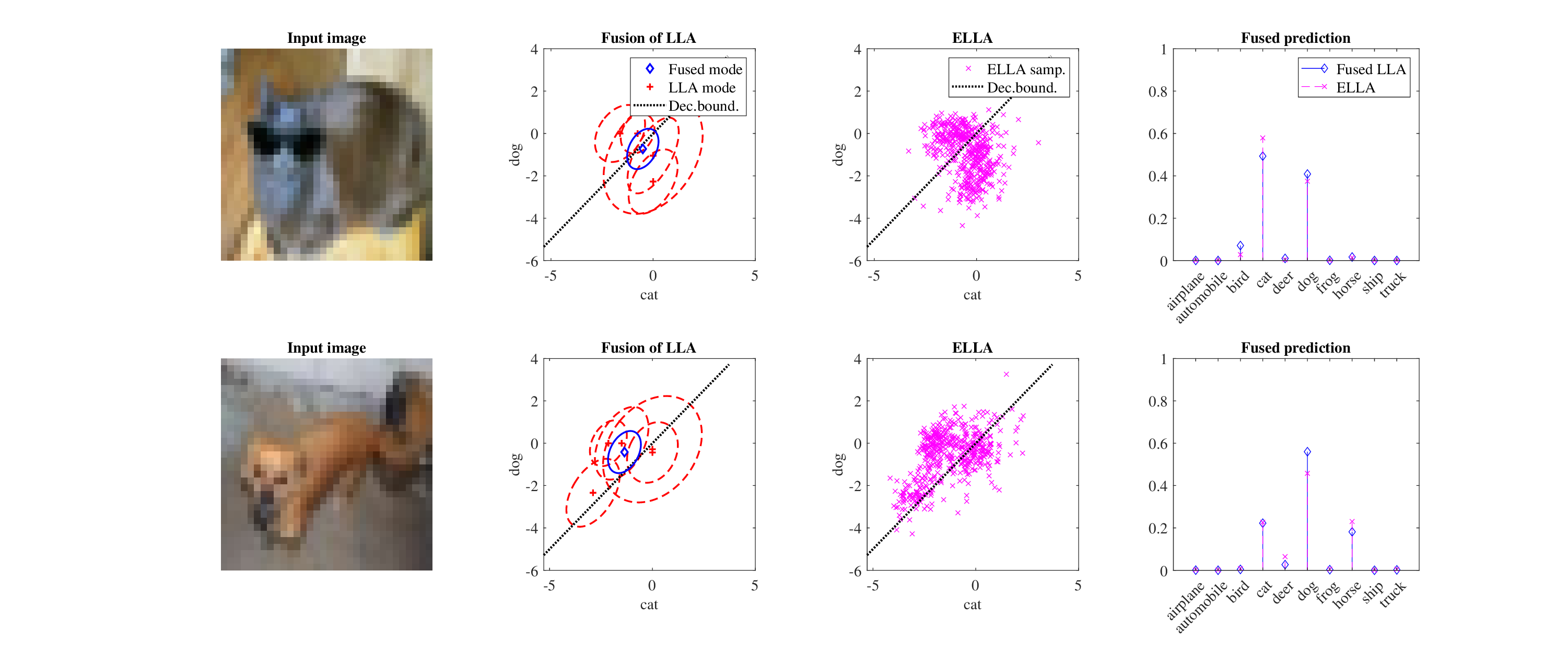}
	\caption{ The images visualize the distribution for the fusion of prediction from multiple \protect{\lla} and the distribution generated by an \protect{\ella}. Here, the visualized using images from \protect{\cfarTen}.  }
	\label{fig:fusion_cfar10}
\end{figure}

\begin{table}[tb!]
	\centering
	\caption{ The computed performance measures for the two datasets. The arrows indicate whether a high or low value is preferable. For a description of measures, see \cref{sec:validation_fm}. }
	\begin{tabular}{p{28mm} | c c c c }	
		\hline
		\multicolumn{1}{c|}{} & \multicolumn{4}{c}{\mnist} \\
		Method & acc. $\uparrow$ & \Loglike $(10^3)$ $\uparrow$ & Brier score $\downarrow$ & \ece $\downarrow$  \\ [0.5ex]
		\hline 
		Temp. sc. \cite{guo2017calibration} & 92$\%$& 7.83 & 0.123 & 2.85   \\
		Deep ensemble \cite{lakshminarayanan2016simple} & \textbf{95$\%$} & 7.94 &0.080 & 2.29\\
		\lla  \cite{malmstrom2023uncertainty} & 92$\%$& 7.81 & 0.125 & 1.33 \\
		Fusion of \lla{s}, \cref{sec:fusion} & 94$\%$& 7.89 & 0.103 & 1.82 \\
		\ella $w_{lc}=1/LC$, \cref{sec:ensablemLinearization} & 93$\%$ & \textbf{8.37} & \textbf{0.071} & \textbf{1.06}  \\
		[1ex]
		\hline
		\hline
		\multicolumn{1}{c|}{} & \multicolumn{4}{c}{\cfarTen}\\
		Method & acc. $\uparrow$ & \Loglike $(10^3)$ $\uparrow$ & Brier score $\downarrow$ & \ece $\downarrow$ \\ [0.5ex]
		\hline 
		Temp. sc. \cite{guo2017calibration}  & 82$\%$ & 7.45 & 0.246 & 1.42  \\
		Deep ensemble \cite{lakshminarayanan2016simple} & 86$\%$ & 7.87 & 0.197 & 1.19\\
		\lla  \cite{malmstrom2023uncertainty} & 82$\%$ & 7.74 & 0.255 & 1.08\\ [0.5ex]
		Fusion of \lla{s}, \cref{sec:fusion} &  86$\%$ &8.04 &0.197 & 0.67\\
		\ella $w_{lc}=1/LC$, \cref{sec:ensablemLinearization}& \textbf{87$\%$} & \textbf{8.20} & \textbf{0.191} & \textbf{0.63} \\
		[1ex]
	\end{tabular}
	\label{table:performance}
\end{table}

\subsection{Receiver operating characteristic, precision, and recall }
The different methods of quantified uncertainty in the prediction can be evaluated on how well out-of-distribution samples can be detected. How well can the prediction of an input that the model has been trained on, in-distribution, be distinguished from an entirely different input, i.e., out-of-distribution examples. Three widely used measures to measure how well out-of-distribution samples are detected are probability of detection, $P_D$ (also known as sensitivity or recall $P_R$) probability of false alarm $P_{FA}$ and precision $P_P$ \cite{Kay1998, Goodfellow2016}. Assume that the $\mathcal{H}_0$ hypothesis is that the samples are out-of-distribution samples and that under the $\mathcal{H}_1$ hypothesis, the samples are in-distribution samples. Then, defining the three quantities as       
\begin{subequations} \label{eq:decteionMeasure}
    \begin{align}
    P_{D} = P_{R} &=  Pr\{ \text{detect } \mathcal{H}_1 | \mathcal{H}_1 \},  \\
    P_{FA} &= Pr\{ \text{detect } \mathcal{H}_1 | \mathcal{H}_0 \}, \\
    P_{P} &= Pr\{   \mathcal{H}_1 | \text{detect }  \mathcal{H}_1 \}. 
    \end{align}
\end{subequations}
The three measures in \eqref{eq:decteionMeasure} will change depending on a set threshold. The receiver operating characteristic (\roc) curve is created by plotting the $P_D$ against the $P_{FA}$ for different threshold values. Similarly, the precision-recall curve is created by plotting the precision against the recall for different threshold values. To evaluate how well different methods are at detecting out-of-distribution samples from in-distribution samples, one often refers to \auroc and \aupr. The \auroc and the \aupr are between zero and one, and the value is one for a perfect detector.

\subsection{Temperature scaling } \label{sec:temp_scaling_fm}
The output of the softmax function can be used as a point estimate of \pmf. However, this does not consider the uncertainty in the parameters, and this approach is known not accurately to reflect the uncertainty in the prediction, i.e., the uncertainty needs to be calibrated \cite{guo2017calibration,vaicenavicius2019evaluating}. It is often overconfident and underestimates the uncertainty. Hence assigning small probability mass to unlikely classes. 

A common approach used to calibrate the estimated \pmf is the so-called temperature scaling \cite{guo2017calibration}. In temperature scaling, $g(x|\theta)$ is scaled by a scalar quantity $T$ before the normalization by the softmax operator. With a slight abuse of notation, introduce
\begin{equation}\label{eq:temp_scaling_fm}
\hat{p}^T_{lc}(y^\star) \triangleq f(x^\star_l|\hat{\theta}_N^c,T)=\text{softmax}\left(g(x^\star|\hat{\theta}_N^c)/T\right).
\end{equation}  
The scaling parameter $T$ is found after training on the model using validation data. The subscript $T$ is used for the estimate of the \pmf when temperature scaling is used.

%

\begin{figure}[bt!]
	\centering
    \subfloat[Very early.][ \label{fig:ood_temperature_fm} Temp. sc.  ]{\includegraphics[trim={0cm 0cm 0cm 0cm} ,clip,width=0.18\textwidth]{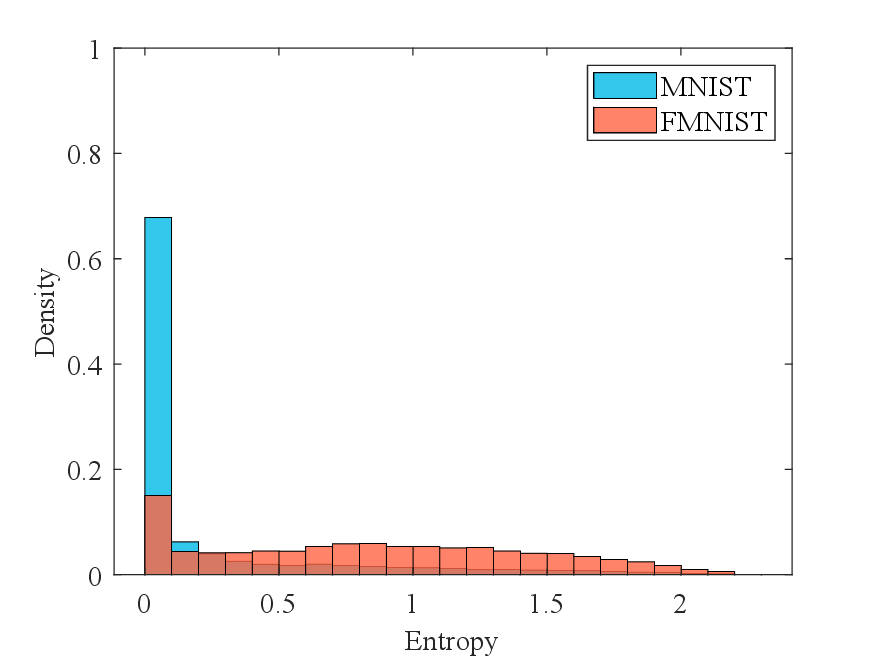}}
    \qquad
    \subfloat[Early.][  \label{fig:ood_ensemble_fm} Deep ensemble.  ]{  		\includegraphics[trim={0cm 0cm 0cm 0cm} ,clip,width=0.18\textwidth]{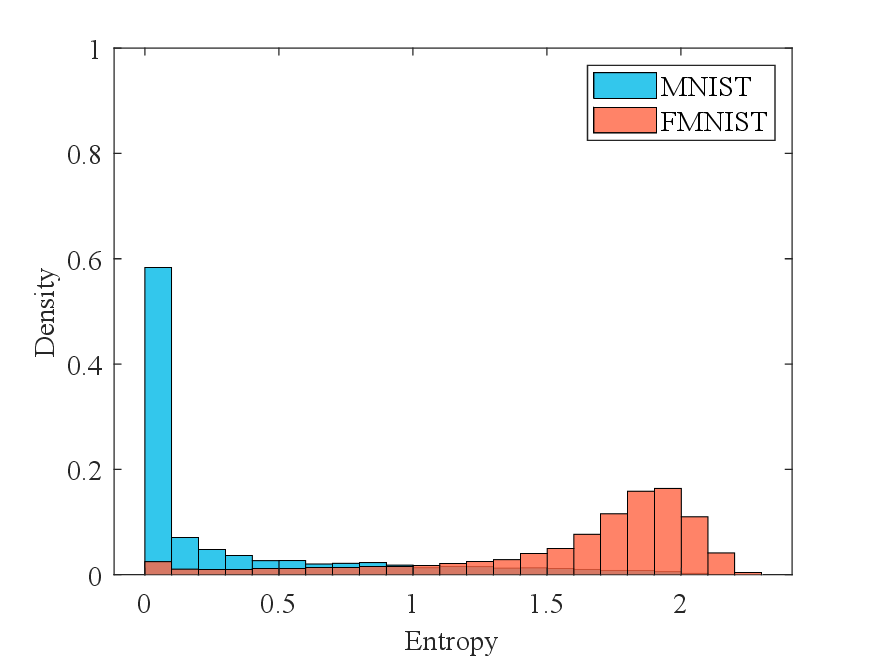}
	}
    \qquad
    \subfloat[Early.][  \label{fig:ood_lla_fm} \protect{\lla.}  ]{  		\includegraphics[trim={0cm 0cm 0cm 0cm} ,clip,width=0.18\textwidth]{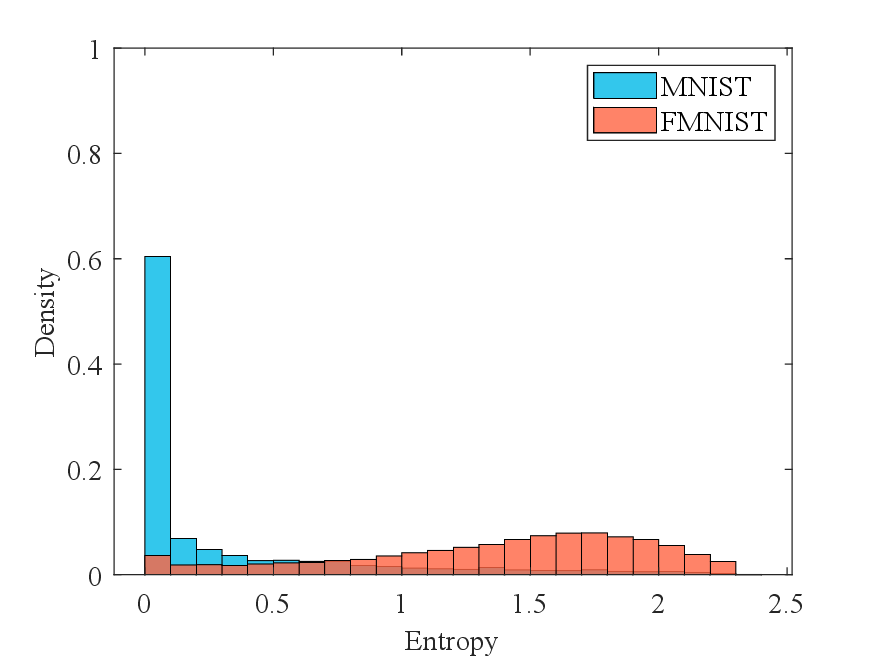} 
	}
	\qquad
	\subfloat[Early.][  \label{fig:ood_fusion_fm} Fusion of \protect{\lla{s}.}   ]{  			\includegraphics[trim={0cm 0cm 0cm 0cm} ,clip,width=0.18\textwidth]{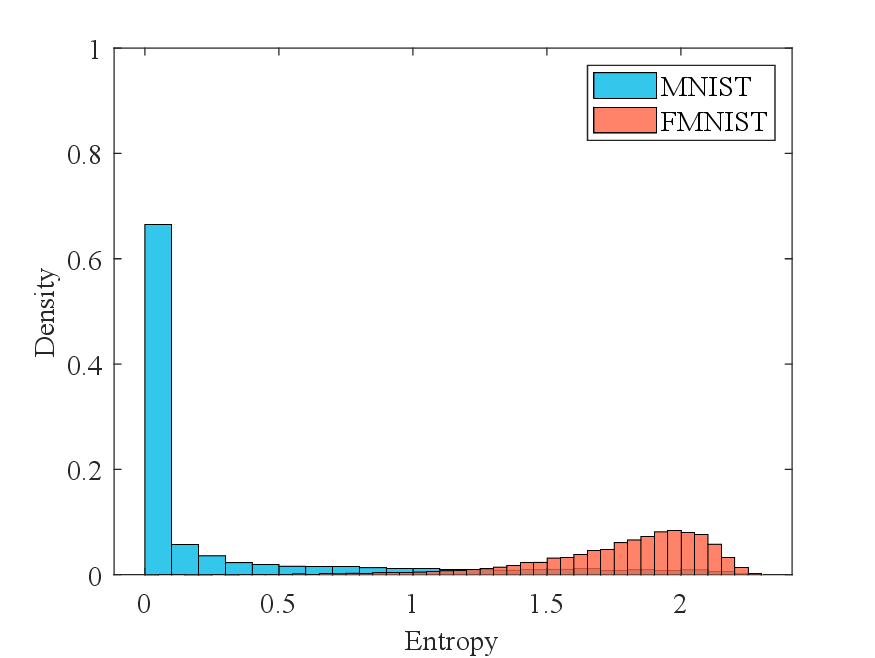}

	}
    \qquad
    \subfloat[Early.][  	\label{fig:ood_ella_fm} \protect{\ella.}  ]{  			\includegraphics[trim={0cm 0cm 0cm 0cm} ,clip,width=0.18\textwidth]{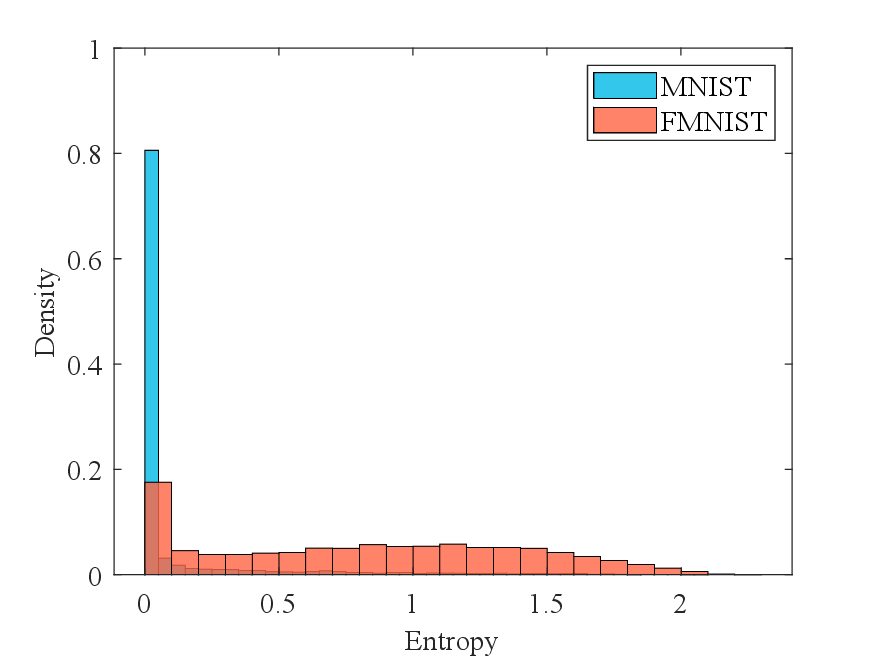}
	}
    \\
    \caption{  Distribution for the predicted entropy \eqref{eq:entropy_fm} for in-distribution \protect{(\mnist)} and out-of-distribution  \protect{(\fmnist)} samples. Here, five different methods to create the estimated conditional \protect{\pmf} are investigated.}
    \label{fig:ood_fm} 
\end{figure}


\begin{table}[tb!]
	\centering
	\caption{ Performance measure for out-of-distribution detection. The arrows indicate whether a high or low value is preferable. See  \cref{sec:validation_fm} for descriptions of the measures. }
	\begin{tabular}{p{35mm} | c c c c }
		\hline
		\multicolumn{1}{c|}{} & \multicolumn{3}{c}{\mnist (in) \fmnist (out) } \\
		Method &  $\Delta_E $ $(10^3)$ $\downarrow$ & \auroc $\uparrow$ & \aupr $\uparrow$ \\ [0.5ex]
		\hline
		Temp. sc. \cite{guo2017calibration} & -6.36 & 0.82 & 0.70  \\
		Deep ensemble \cite{lakshminarayanan2016simple} & -13.12 & 0.94 & 0.91\\
		\lla  \cite{malmstrom2023uncertainty}& -11.23 & 0.91 & 0.91 \\
		Fusion of \lla{s}, \cref{sec:fusion}  & \textbf{-13.83}& \textbf{0.95}  & \textbf{0.96}\\
		\ella $w_{lc}=1/LC$, \cref{sec:ensablemLinearization} & -5.42 & 0.87& 0.83 \\
		[1ex]
	\end{tabular}
	\label{table:2ood_fm}
\end{table}

\begin{figure*}[tb!]
	\centering
  \includegraphics[trim={4cm 0cm 4cm 0cm},clip,width=1.0\textwidth]{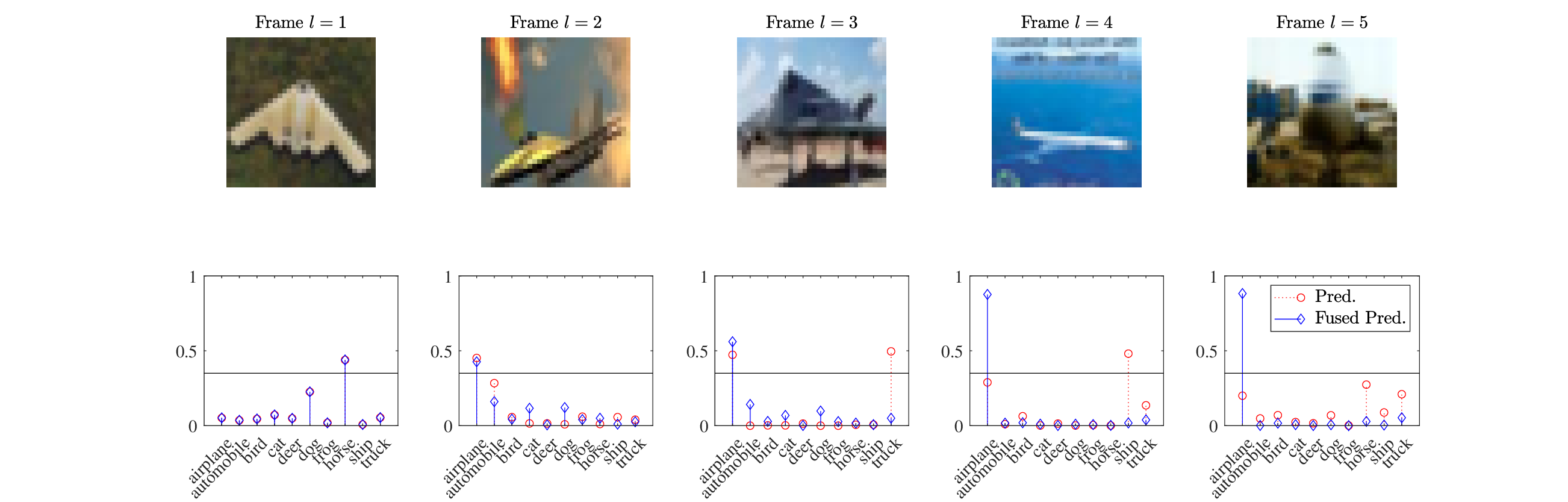}
	\caption{ The image illustrates the fusion of a sequence of images from the \protect{\cfarTen} dataset. Here, using the same \protect{\nn} to predict the class for all the images in the sequence and using  \eqref{eq:fusion_seq} to fuse the prediction. Hence, the cross-correlation between the predictions is taken into consideration.   }
	\label{fig:seqfusion_cfar10}
\end{figure*}

\section{Experiment study}
The experiments in this section can be divided into two parts. The first part investigates the fusion of multiple classifiers that classify the output for the same input. Here, an ensemble of \nn{s} are trained on the two classical data sets \mnist \cite{lecun1998mnist} and \cfarTen \cite{krizhevsky2009learning}.
The \nn trained on \mnist will be evaluated on Fashion \mnist (\fmnist) \cite{xiao2017fashion} to investigate how to detect out-of-distribution examples. This part investigates the fusion strategy described in \cref{sec:fusion} and \ella described in \cref{sec:ensablemLinearization}.   
In the second part, we are simulating the setup where a sequence of images known to belong to the same class by giving a classifier a sequence of images from the \cfarTen data set and data from a camera trap that captures images of Swedish carnivores.

For the three different classification tasks, i.e., \mnist, \cfarTen, and camera trap images of Swedish carnivores, different structures of the \nn{s} were used. For the \mnist task, a five-layer fully connected \nn was used; for the \cfarTen task, using a LeNet5 \cite{lecun1998gradient} inspired structure; and for the Swedish carnivores, using an AlexNet \cite{krizhevsky2012imagenet} inspired structure.     
Here, using the \adam optimizer \cite{Kingma2015} with the standard setting to estimate parameters $\theta$ for all the tasks. The training was done for three epochs for the \mnist task, ten epochs for the \cfarTen, and six epochs for the Swedish carnivores. 

This paper is investigating five different methods to quantify the uncertainty, namely 
\begin{enumerate}[label=(\roman*)]
    \item Temperature scaling \cite{guo2017calibration}, see \cref{sec:temp_scaling_fm}.
    \item Deep ensemble \cite{lakshminarayanan2016simple}
    \item One single \lla, \cite{malmstrom2023uncertainty}.
    \item Fusion of the prediction of an ensemble of \lla, see \cref{sec:fusion}.
    \item An \ella where every mode is equally weighted, i.e., \eqref{eq:multimodality} with $w_{lc} = 1/LC$.
\end{enumerate}
For the deep ensemble, the size of the ensemble was ten for the \mnist task and five for the  \cfarTen task. This was also the number of \nn{s} trained when fusing the prediction from multiple \lla{s} and the number of modes in the \ella. 
\subsection{Fusion of multiple measurements }
Having access to predictions from multiple independently trained \nn{s}, each associated with an \lla, it is possible to fuse their prediction to better understand the predicted output of a given input. 
\cref{table:performance} shows the result of the experiment done on \mnist and \cfarTen .
For the \cfarTen data set, all the measures are improving for the fusion of \lla{s} compared to only using the temperature scaling, deep ensemble, and a single \lla. In particular, the \ece decreases. However, for the \mnist data set, the \ece does not decrease for the fusion of \lla{s} compared to only using a single \lla. 
That the fusion of \lla{s} does not decrease the \ece can be a consequence of that some information about the distribution goes missing in the fusion stage. 
An example of the fusion of the prediction can be seen in the second column in \cref{fig:fusion_cfar10}. Here, the mean and standard deviation for the different modes are shown in red, and the fused prediction is in blue. The fourth column shows the estimated \pmf for the fused prediction. 

In \cref{fig:ood_fm}, the distribution for the predicted entropy for images from both images the \nn is trained on (in-distribution) and images the \nn is not trained on (out-of-distribution). The in distribution is \mnist, i.e., images of numbers, and the out of distribution is \fmnist, i.e., images of clothing items. Compared to the other methods, fusing the prediction from multiple \lla{s} seen in \cref{fig:ood_fusion_fm} provides a more evident visual difference between the in and out distribution images. This claim is also supported in \cref{table:2ood_fm} where the \auroc and \aupr are better for the fused predictions. The difference in the predicted entropy for in and out of distribution samples for the fused predictions is also significant; see \cref{table:2ood_fm}.

\subsection{Sequential fusion}
In \cref{fig:seqfusion_cfar10}, a sequence of six images of aircraft from \cfarTen is shown. Here, classifying all the images in the sequence by the same \nn. Hence, the predictions are correlated according to \eqref{eq:fusion_seq}. Assuming that all the images in the sequence belong to the same class, this information can be used to improve the prediction accuracy from images later in the sequence. For example, even though the prediction of the fourth image is uncertain,  thanks to the information from the previous images, the fused prediction indicates that it is likely that the object in image is an aircraft.

\begin{figure}[tb!]
	\centering
\includegraphics[trim={5cm 0cm 3cm 0cm},clip,width=0.45\textwidth]{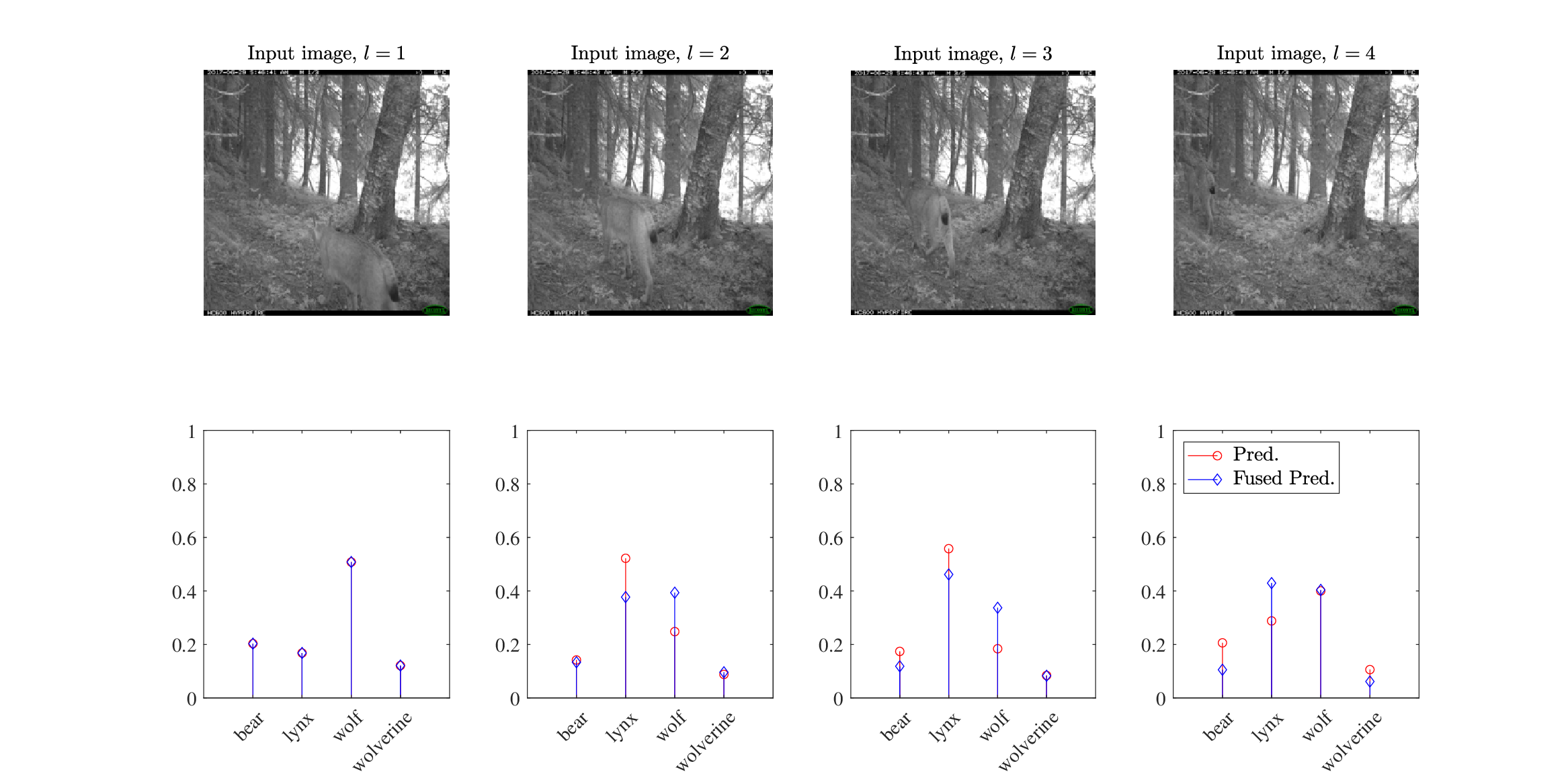} 
	\caption{  The figure illustrates the fusion of the predicted class from an image sequence of a lynx. Here, using the same \protect{\nn} to predict the object's class in all the images in the sequence, where \eqref{eq:fusion_seq} is used to fuse the prediction over the sequence. }
	\label{fig:seqfusion_lynx}
\end{figure}

\cref{fig:seqfusion_lynx} illustrates an experiment where the prediction of multiple images from a camera trap. Here, a lynx is passing the camera, and even though the last image in the sequence is classified as a wolf, by using the information from the previous images, the fused prediction still indicates that it is a lynx in the image.

\subsection{Multimodal model}
It is possible to have a multimodal representation of the prediction distribution with access to an \ella. To visualize the multimodality, one can draw independent samples from the different modes created by the different \nn in the ensemble.  \cref{fig:fusion_cfar10} is visualizing  this for two images from the \cfarTen dataset. The third column shows a representation of the distribution before the normalization. Here, independent \mc-samples are taken from the different modes to visualize the distribution.

In \cref{table:performance}, it can be seen that inclusion of the multimodality improves calibration of the predicted uncertainty, i.e., leads to lower a \ece. This while still maintaining high accuracy and high \Loglike. The use of the \ella to detect out-of-distribution samples can be seen in \cref{fig:ood_ella_fm} and \cref{table:2ood_fm}. Here, it is shown that multimodality improves the detection compared to temperature scaling, but the improvements could be better when fusing the prediction from multiple \lla{s}.

\section{Summary and conclusion}
This paper suggests two methods to estimate the conditional \pmf given a sequence of classifications when one can estimate the uncertainty in the prediction from an \nn. The first method is the fusion of the predictions, where both the information from the previous and the cross-correlation between the predictions are considered. 
The suggested method to include uncertainty in the prediction from an \nn used in the fusion strategy is called \lla. 
 The second method extends the \lla to represent multimodality in the predicted distribution. This extension is referred to as \ella, where it combines a deep ensemble and \lla. 

The fusion strategies and the proposed extension are evaluated on classification tasks with images from the classical data sets \mnist and \cfarTen and images from camera traps collected in the Swedish forests. It also investigates how to use the methods to detect out-of-distribution samples. Here, \mnist is used as in-distribution and \fmnist as out-of-distribution samples. 
Sequential fusion of the predictions is shown to increase the robustness of the prediction. Both when all predictions in the sequence are uncertain and only one prediction is confident but incorrect. 
The results in terms of \ece show that the multimodal representation using the extended \ella method improves the performance compared to previous work. 
In most cases, the fusion of multiple predictions improved the performance. 
Furthermore, when detecting out-of-distribution samples, the fusion of predictions from multiple \nn leads to improvement in performance. This improvement is in terms of increased \auroc and \aupr and the difference in predicted entropy. 
Here, the performance using \ella could be improved. This indicates that some modes might be more critical in detecting out-of-distribution samples. Hence, methods to give the modes different weights are a possible direction for future research.

%

\addtolength{\textheight}{-4.2cm}   



%


\bibliography{IEEEabrv,mybibref5G}

\end{document}